\documentclass[twocolumn]{aastex631}

\newcommand{\rs}{R$_\odot$} % solar radius
\def\submitted#1{}
\journalinfo{Accepted for publication in ApJ on February 20, 2025}

 \usepackage{savesym}       % Workaround to use siunitx
    \savesymbol{tablenum}
    \usepackage{siunitx}
        \restoresymbol{SIX}{tablenum}   % resolve tablenum conflict
 \usepackage{wrapfig} % Allows wrapping text around tables and figures
 \usepackage{subfigure}
\usepackage{xcolor}
\usepackage{amsmath} 
\begin{document}

\title{Determining the polarisation of a coronal standing kink oscillation using spectral imaging techniques with CoMP
\footnote{The Coronal Multi-channel Polarimeter}
}

\author[0000-0003-3306-4978]{T.~J.~Duckenfield}
\affiliation{Astrophysics Research Centre, \\
School of Mathematics and Physics, \\ 
Queen’s University Belfast, \\
Belfast, BT7 1NN, UK}
\affiliation{Department of Mathematics, Physics and Electrical Engineering, \\
Northumbria University, \\
Newcastle Upon Tyne, \\
NE1 8ST, UK}

\correspondingauthor{T.~J.~Duckenfield}
\email{t.duckenfield@qub.ac.uk}

\author[0000-0002-9155-8039]{D.B.~Jess}
\affiliation{Astrophysics Research Centre, \\
School of Mathematics and Physics, \\ 
Queen’s University Belfast, \\
Belfast, BT7 1NN, UK}
\affiliation{Department of Physics and Astronomy, \\ California State University Northridge, \\
USA}

\author[0000-0001-5678-9002]{R.~J.~Morton}
\affiliation{Department of Mathematics, Physics and Electrical Engineering, \\
Northumbria University, \\
Newcastle Upon Tyne, \\
NE1 8ST, UK}

\author[0000-0002-7711-5397]{S.~Jafarzadeh}
\affiliation{Astrophysics Research Centre, \\
School of Mathematics and Physics, \\ 
Queen’s University Belfast, \\
Belfast, BT7 1NN, UK}

\begin{abstract}

%Context
Coronal oscillations offer insight into energy transport and driving in the solar atmosphere. Knowing its polarisation state helps constrain a wave's displacement and velocity amplitude, improving estimates of wave energy flux and deposition rate.
We demonstrate a method to combine imaging and spectral data to infer the polarisation of a coronal loop's standing kink wave, without the need for multiple instruments or multiple lines of sight.
%Methods
We use the unique capabilities of the Coronal Multi-channel Polarimeter (CoMP) to observe the standing kink mode of an off-limb coronal loop perturbed by an eruption.
The full off-disk corona is observed using the \qty{1074}{\nm} Fe~{\sc{xiii}} spectral line, providing Doppler velocity, intensity and line width.
%Results
By tracking the oscillatory motion of a loop apex in a time-distance map, we extract the line-of-sight (Doppler) velocity of the inhomogeneity as it sways and compare it with the derivative of its plane-of-sky displacement.
This analysis provides the loop's velocity in two perpendicular planes as it oscillates with a period of $8.9^{+0.5}_{-0.5}$~minutes.
Through detailed analysis of the phase relation between the transverse velocities we infer the kink oscillation to be horizontally polarised, 
oscillating in a plane tilted $-13.6^{+2.9}_{-3.0} \unit{\degree}$ away from the plane of sky.
The line widths show a periodic enhancement during the kink oscillation, exhibiting both the kink period and its double.
This study is the first to combine direct imaging and spectral data to infer the polarisation of a coronal loop oscillation from a single viewpoint.

\end{abstract}

\keywords{magnetohydrodynamics (MHD) -- Sun: corona -- Sun: magnetic fields -- Sun: oscillations -- waves}

%%%%%%%%%%%%%%%%%%%%%%%%%%
\section{Introduction}
\label{sec:intro}
%%%%%%%%%%%%%%%%%%%%%%%%%

The solar corona is a highly dynamic environment, hosting a wide range of wave phenomena which can be used to probe the local plasma conditions \citep{Nakariakov2024_review}. 
Among these, magnetohydrodynamic (MHD) kink oscillations are particularly useful, due to their widespread occurrence and well-established ability to probe the magnetic and density structuring of their host inhomogeneities across the solar atmosphere, such as coronal loops, %\citep{Verwichte2009_polarisation}
prominences, %\citep{Mazumder2020}
fibrils, and spicules \citep[see the recent reviews of][]{2021SSRv..217...73N, 2023LRSP...20....1J}. 

These collective transverse motions of a (typically density-enhanced) cylinder are characterised by an azimuthal wavenumber, $m$, equal to 1 \citep{Ruderman2009_KinkReview}. 
The phase speed of the wave is equal to the kink speed, $C_K$, which is a density-weighted average of the Alfv\'en speed across the inhomogeneity. %loop. 
Coronal loops often exhibit kink oscillations, and since there is wave reflection at the loop footpoints at the much denser transition region and chromosphere, a longitudinal standing mode (and its higher harmonics) is regularly formed. 
The period of the $n$-th longitudinal harmonic is determined by the loop length, $L$, following the relationship 
$P=2L/(nC_K)$ \citep{Nakariakov2024_review}. The excitation of these longitudinal harmonics is governed by the characteristics of the driver \citep{Zimovets2015}.
However, higher harmonics damp more rapidly %\citep{Ruderman2009_KinkReview} %\citep{Amiri2021}
and require greater spatial and temporal resolution to detect, resulting in most observations of standing kink oscillations being related to the fundamental mode.
Kink modes may be damped extremely quickly (in a matter of a few periods) through the process of resonant absorption \citep{Goossens2002,RudermanRoberts2002,Pascoe2016_modecoupling_pt1,Guo2020_elliptical1,Morton2021_weakdamping}. 
The bulk motion of the plasma within the loop cross-section couples to localised Alfv\'enic motions, at a particular radius where $C_K = v_A(r)$, since the cross-sectional inhomogeneity leads to a radially varying Alfv\'en speed. 
These internal motions remove energy from the kink oscillation and lead to strong damping, usually with a quality factor (ratio of damping time to period) of $\sim 1-10$. 
%\subsection{Seismology}    
The connection between standing kink oscillations and the local plasma parameters has been extensively exploited to diagnose the coronal plasma via coronal seismology \citep{Nakariakov2024_review}.  
%\subsection{Polarisations}
Transverse oscillations of coronal loops can have two linear polarisations -- a vertical polarisation indicates motion in a plane perpendicular to the solar surface, while horizontal polarisation corresponds to displacement parallel to the surface.
In addition, transverse oscillations of coronal loops can exhibit elliptical or circular polarisation indicative of coupled displacements in both vertical and horizontal directions, resulting in a rotational or helical motion of the loop's axis.
The damping by resonant absorption is independent of whether the wave is linearly, circularly, or elliptically polarised \citep{Magyar2021_circpol}. 
However, determining the wave polarisation allows more accurate constraints on its true displacements and velocity amplitudes.
This directly improves estimates of wave energy fluxes and deposition rates, which are of crucial importance for assessing the contribution of waves to coronal heating, and the transport of energy through the atmosphere.

% Context figure
\begin{figure}[t]
    \begin{center}
    %\begin{interactive}{animation}{fig1.mp4}
    \resizebox{\hsize}{!}{\includegraphics{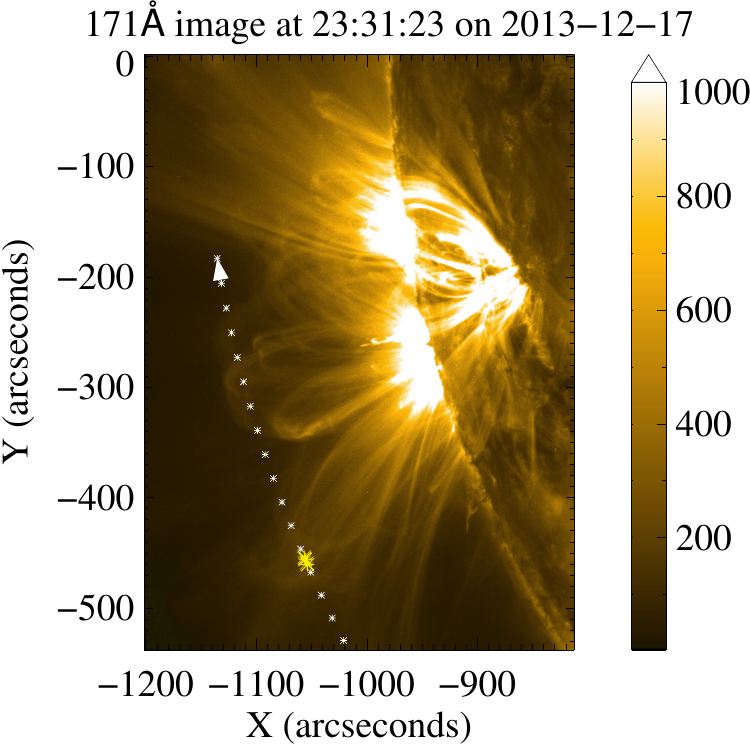}}
    %\end{interactive}
    \end{center}
    \caption{Active Region 11925 observed by the Solar Dynamics Observatory's Atmospheric Imaging Assembly in the \qty{171}{\angstrom} channel.
    The white dashed line represents a slit positioned parallel to the solar surface at a radius of 1.18 \rs.
    The yellow points trace the measured displacement of the oscillating feature studied.
    }
    \label{fig:context_171}
\end{figure}
In coronal loop observations, the majority of kink modes are thought to be in the horizontal polarisation mode, excited by eruptive events \citep[e.g.,][]{2011A&A...534A..13B, Zimovets2015}. 
The prevalence of horizontal modes is presumably because a general excitation mechanism, such as an eruption, is more likely to impact the loop from the side, rather than below/above. 
Indeed, the recent work by \citet{Zhong2023_polarisation} determined the horizontal (or weakly oblique) polarisation of a large scale kink mode utilising the perspective of multiple instruments. 
Vertically polarised modes have been detected in individual cases,for example in \citet{Verwichte2009_polarisation} and \citet{Aschwanden2011_polarisation}. 
However, the need for multiple perspectives to determine polarisation limits studies, as it requires the same oscillation event to be seen by two different instruments.
Moreover, without a priori knowledge of the line-of-sight angle, there is no unambiguous signature of the polarisation state in the forward-modelled Doppler velocities from the induced internal flows \citep{Goossens2014}. %Only through combining the plane-of-sky velocities with either the Doppler velocities or another point of view entirely, can the polarisation state may be found. 

In this work, we demonstrate a novel method for inferring the polarisation of a specific loop oscillation by analysing its velocity phase portrait, combining observations of the loop's transverse motion in the plane of the sky with the corresponding line-of-sight Doppler velocities. 
Unlike traditional approaches that require stereoscopic observations from multiple viewpoints, this technique enables the characterisation of an oscillation's polarisation using data from a single instrument.

\section{Observations} 
\label{sec:observations}

\subsection{Coronal Multi-Channel Polarimeter (CoMP)}
\label{subsec:CoMP}
The Coronal Multi-Channel Polarimeter instrument \citep[CoMP;][]{Tomczyk2008_CoMP} is a ground-based instrument designed to observe the solar corona, operating from the Mauna Loa Solar Observatory in Hawaii from 2013 -- 2018.
At a sub-minute cadence occulted images are taken of the solar corona, and at each pixel a spectrum is extracted with particular targeting of the `forbidden' Fe~{\sc{xiii}} 1074.7~nm and 1079.8~nm near-infrared spectral lines. 
These spectral profiles are fitted with Gaussian functions to determine the Doppler velocity and line intensity, and the intensity ratio can be used to map the coronal electron number density \citep[which assuming overall charge neutrality can indicate plasma density, c.f.][]{Yang2020_globalB}.

% \subsection{Underestimation of CoMP Doppler velocities}
% \label{subsec:underestimation}
While CoMP is a powerful tool for studying solar dynamics, there is a well-documented tendency for spectrometers observing coronal lines to underestimate Doppler velocities.
This underestimation is due to the physics of radiative transfer through optically thin plasmas, in addition to the superposition of out-of-phase oscillations along the line of sight, which can broaden spectral lines and diminish the measured velocity amplitudes \citep{McIntosh2012,Morton2015,Pant2019_darkenergy,DeMoortel2012_multistrand}, especially in the quiet Sun.
Being optically thin, the \qty{1074}{\nm} emission used in these observations is weighted towards the higher density plasma, which in our case is ideal since we are interested in the more dense active region loop.
Additionally, kink oscillations in active regions are coherent over spatial scales larger than CoMP’s resolution \citep{Sharma2023_scales}, allowing the loop to stand out against its surroundings when integrated along the line of sight. 
However, it is important to consider the potential systematic uncertainty in the Doppler velocities. % and their impact on our interpretation of polarisation.
A recent study \citep{Lee2021_CoMP} compared Hinode/EIS data with CoMP~\qty{1074}{\nm} observations and found that whilst the line widths and Doppler velocities are generally correlated, in bright active region structures the EIS Doppler velocities are larger than CoMP's absolute values by a factor of $\sim1.5$. 

% Previous results of CoMP
Previous work using CoMP data have shown an abundance of propagating Alfv\'enic waves propagating both inwards and out of the corona, which appear only weakly damped \citep{Morton2021_weakdamping}, and have a transverse correlation length of \qtyrange{7.6}{9.3}{\mega \metre} associated with supergranulation \citep{Sharma2023_scales}.
Using the propagation speed of these waves, combined with the density estimate from the Fe~{\sc{xiii}} forbidden line ratio, global maps of the coronal magnetic field can be made \citep{Yang2024_uCoMP_longterm}.

\subsection{Event description}
\label{subsec:event}

% Angular CoMP context figure
\begin{figure}[t]
	\begin{center}
        %\begin{interactive}{animation}{fig2.mp4}
        \resizebox{\hsize}{!}{\includegraphics{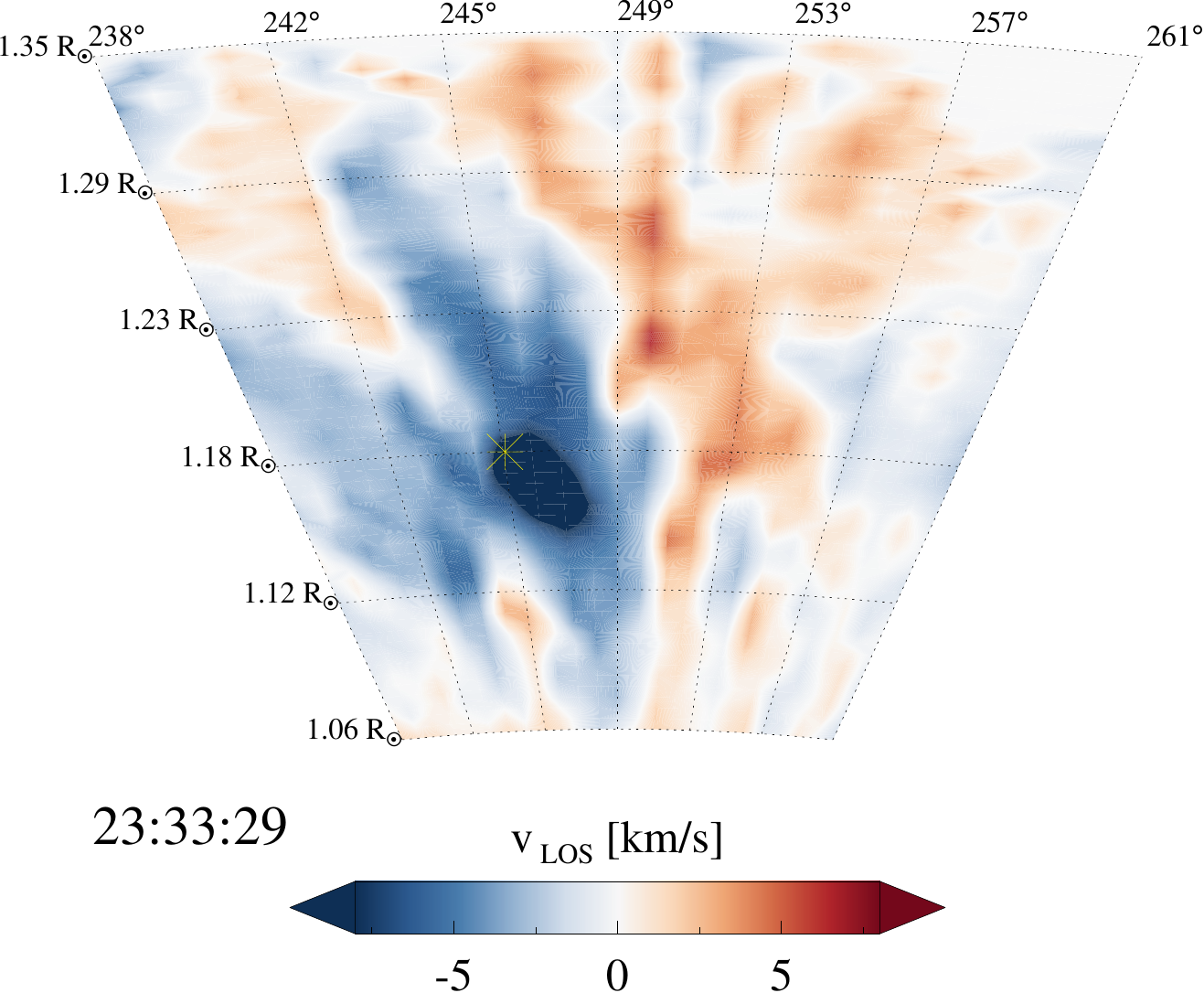}}
        \resizebox{\hsize}{!}{\includegraphics{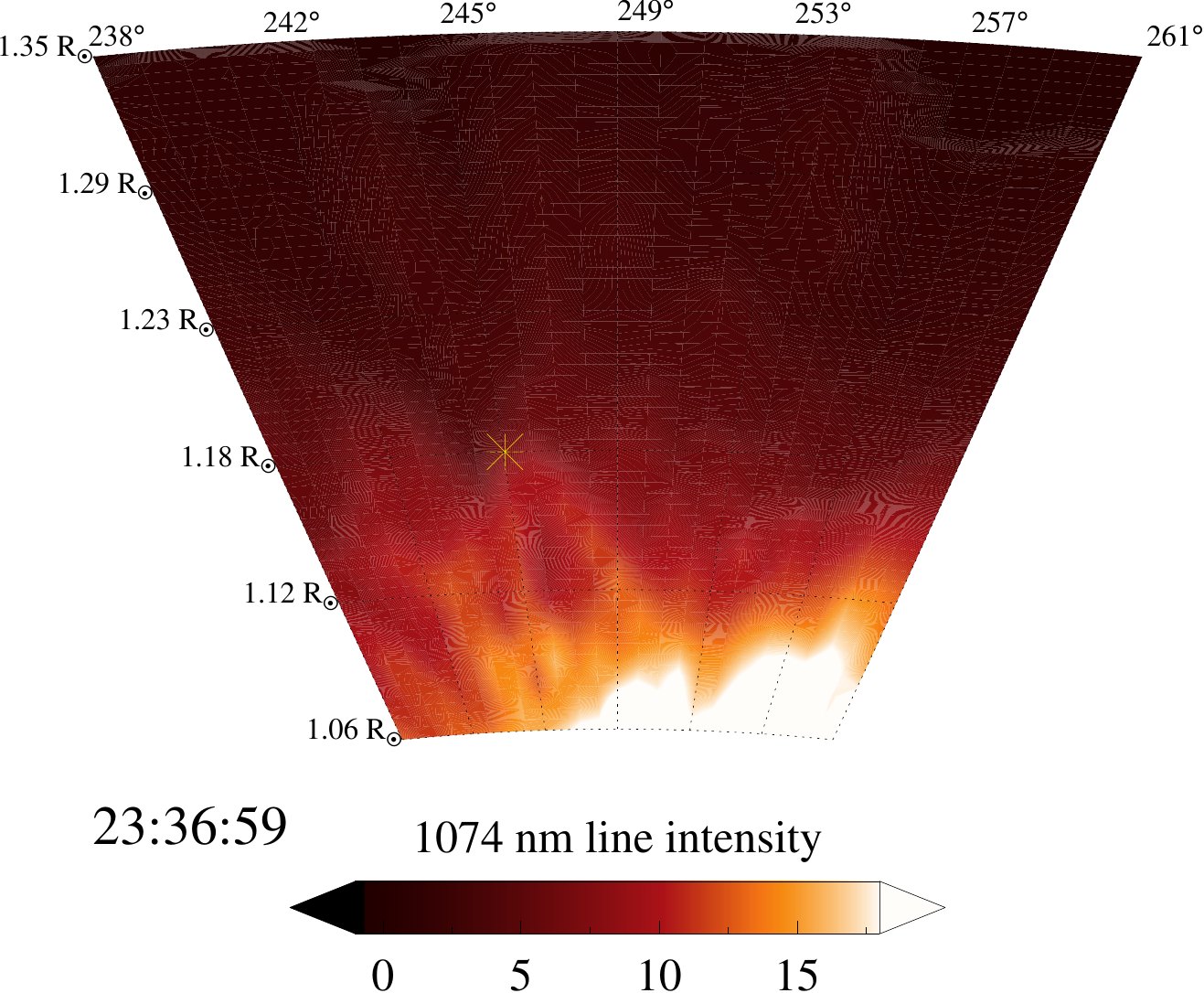}}
        %\end{interactive}
	\end{center}
	\caption{Plots showing Doppler velocity (top) and line intensity (bottom) measured by CoMP as the eruption propagates outwards.
    The angular extent is \qtyrange{220}{270}{\degree} measured clockwise from solar north, and covers a radial range from 1.06 -- 1.35 \rs.
    The top plot shows a blueshifted loop apex (marked with a yellow cross) superimposed on a background of redshifted plasma from the erupting material.
    The bottom plot, separated by approximately 4 minutes (roughly half the oscillation period), reveals the full extent of the loop of interest, including its legs.
	}
	\label{fig:angrs}
\end{figure}
To demonstrate the ability of the CoMP instrument in combining Doppler velocity and imaging, we consider an off-limb large-scale kink oscillation observed on 2013 December 17. 
There is a minor eruption off the eastern solar limb commencing around 23:25, perturbing a bundle of coronal loops (associated with NOAA active region 11925) that causes large amplitude transverse oscillations that last approximately 40 minutes, as seen in SDO/AIA \qty{193}{\angstrom} and \qty{171}{\angstrom} (Figure~\ref{fig:context_171}).
The CoMP instrument was running nearly continuously at 30 second cadence from 23:21 onwards, capturing the eruption in the \qty{1074}{\nm} line only. 
The \qty{1079}{\nm} line information is unavailable, and due to the absence of line intensity ratio measurements, traditional magneto-seismology methods commonly applied with CoMP data \citep[such as in][]{Yang2020_globalB} are not feasible in this case. 
Remarkably, despite this limitation, it remains possible to deduce the polarisation state of the oscillation using only a single line profile, showcasing the robustness of CoMP's capabilities in diagnosing oscillatory phenomena.
From the \qty{1074}{\nm} profile, fitted by a 3 point Gaussian, maps of intensity, Doppler velocity, and line widths are found. 

The region of the corona in the CoMP data which is visibly affected by the eruption is shown in Figure~\ref{fig:angrs}. 
The erupted material appears on this plot as a column of redshifted plasma, with minimal intensity perturbations in this bandpass, moving radially outward and crossing a distance of 1.06 solar radii at approximately \qty{255}{\degree}. 

Whilst the SDO/AIA data depicts a multitude of loops in this region, most of which are perturbed by the eruption, in the CoMP \qty{1074}{\nm} data only one coronal loop is easily discernable in the intensity image foreground. 
At this loop's apex, marked by a yellow point at its equilibrium position ($-1054''$, $-454''$), strong and periodic line-of-sight velocities are observed, also evident in the loop legs to a lesser extent.

For this event, detailed investigation found that the maximum displacement in the plane of the sky aligns with the azimuthal direction. 
In the general case, tracking loop motion along both azimuthal and radial directions within the plane of the sky is ideal to fully capture the loop's dynamics. 
However, in this specific case, radial displacement is negligible. 
Thus, comparing the azimuthal displacement with the line-of-sight velocity provides sufficient information to determine the loop's polarisation state.
%For example, if a loop undergoes significant radial displacement and its plane of oscillation is nearly aligned with the plane of the sky, any circular or elliptical polarisation in the oscillation would manifest as a phase difference between the radial and azimuthal displacements or velocities, rather than between these components and the line-of-sight velocity.

\begin{figure}[ht]
    \gridline{\fig{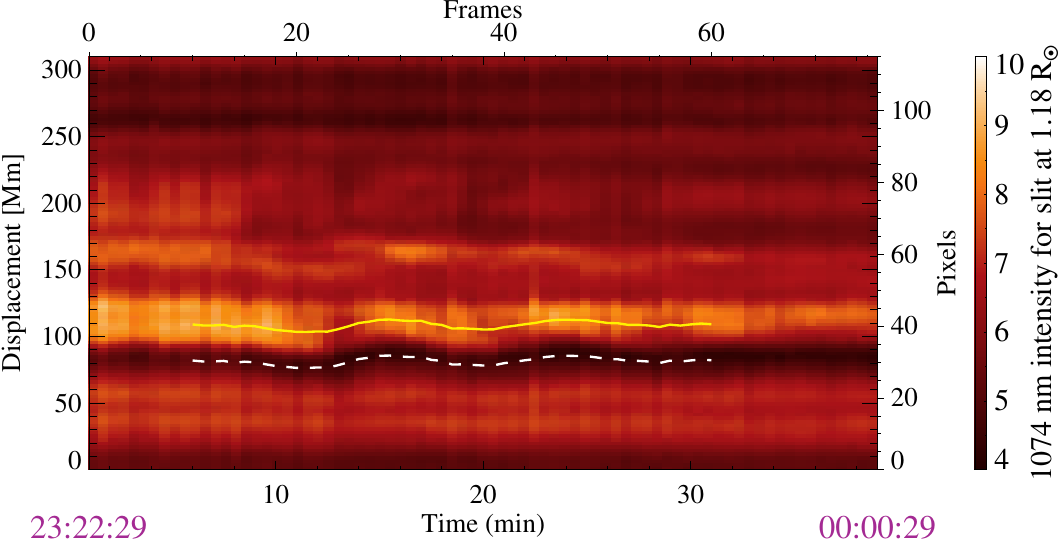}{\linewidth}{(a) Time-distance plot of \qty{1074}{\nm} intensity. The fitted displacement is overlaid in yellow, while the dashed white line represents the fit to the contrasted dark feature prior to the shift.}}
    \gridline{\fig{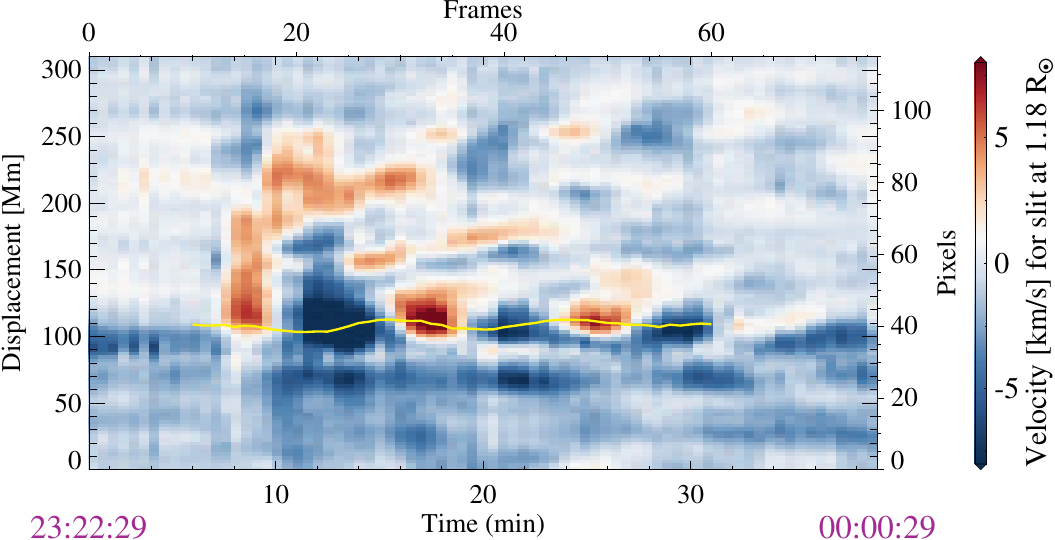}{\linewidth}{(b) Time-distance plot of \qty{1074}{\nm} LOS velocity.}}
    \gridline{\fig{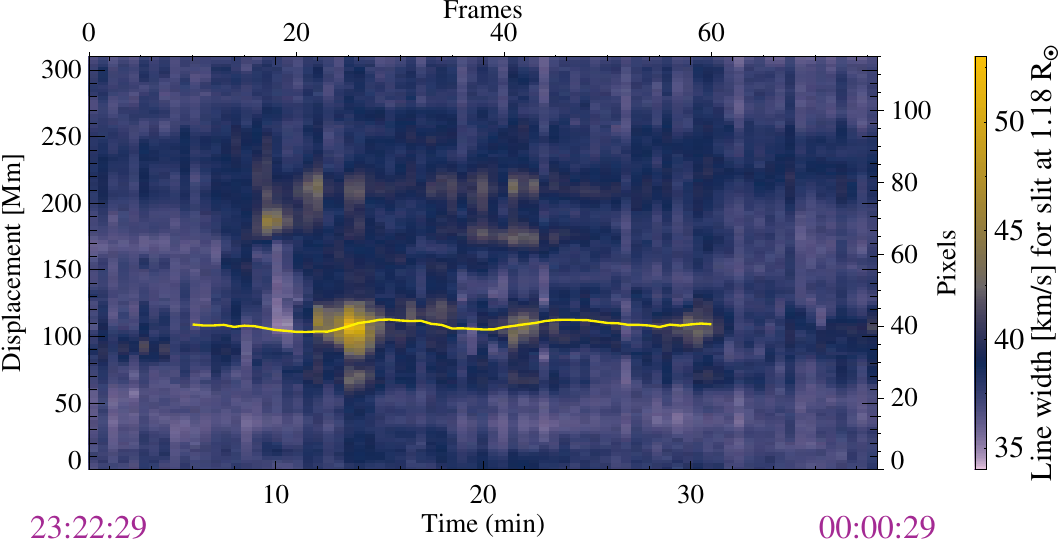}{\linewidth}{(c) Time-distance plot of \qty{1074}{\nm} line width.}}
    \caption{Time distance plots for the slit parallel to the solar surface at 1.18 \rs, i.e. approximately 125.23~Mm above the limb, between angles 238-260 degrees clockwise from solar north.}
    \label{fig:td_maps}
\end{figure}

A series of azimuthal slits approximately 500 arcseconds long and parallel to the solar surface were made from 1.06 -- 1.35 {\rs} in increments of 0.01 \rs, centred at \qty{250}{\degree} from solar north (clockwise) to capture the eruption and subsequent oscillation. 
From these slits time-distance maps were extracted, such as those shown in Figure~\ref{fig:td_maps}. 
The most prominent oscillatory signal in the CoMP data is visible at the loop apex which corresponds to a height of 1.18~{\rs} and angle from solar north of \qty{246}{\degree}.
% Show td-maps

\subsection{Fitting loop displacement}
\label{subsec:fitting}
To compare line-of-sight (LOS) velocity with plane-of-sky (POS) motion, we must precisely map the displacement of a specific inhomogeneity (loop). 
Previous studies have often relied on manual selection of loop positions or simple Gaussian fitting of intensity structures, but these approaches are insufficient for this study, as we are primarily interested in the derivative of displacement. 
Minor inaccuracies in displacement would be amplified in its derivative.
To ensure a precise fit, we use the Solar Bayesian Analysis Toolkit \citep[SoBaT;][]{Anfinogentov2020_SOBAT}, employing Bayesian inference and Markov Chain Monte Carlo (MCMC) sampling to model the intensity profile across the time-distance map with a density enhancement plus background. 
Detailed modelling of line-of-sight integrated density enhancements can be found in \citet{Pascoe2016_modecoupling_pt1}.

\begin{figure}[ht!]
    %\plotone{qcshift.pdf}
    %\begin{interactive}{animation}{fig4.mp4}
    \includegraphics[width=\linewidth]{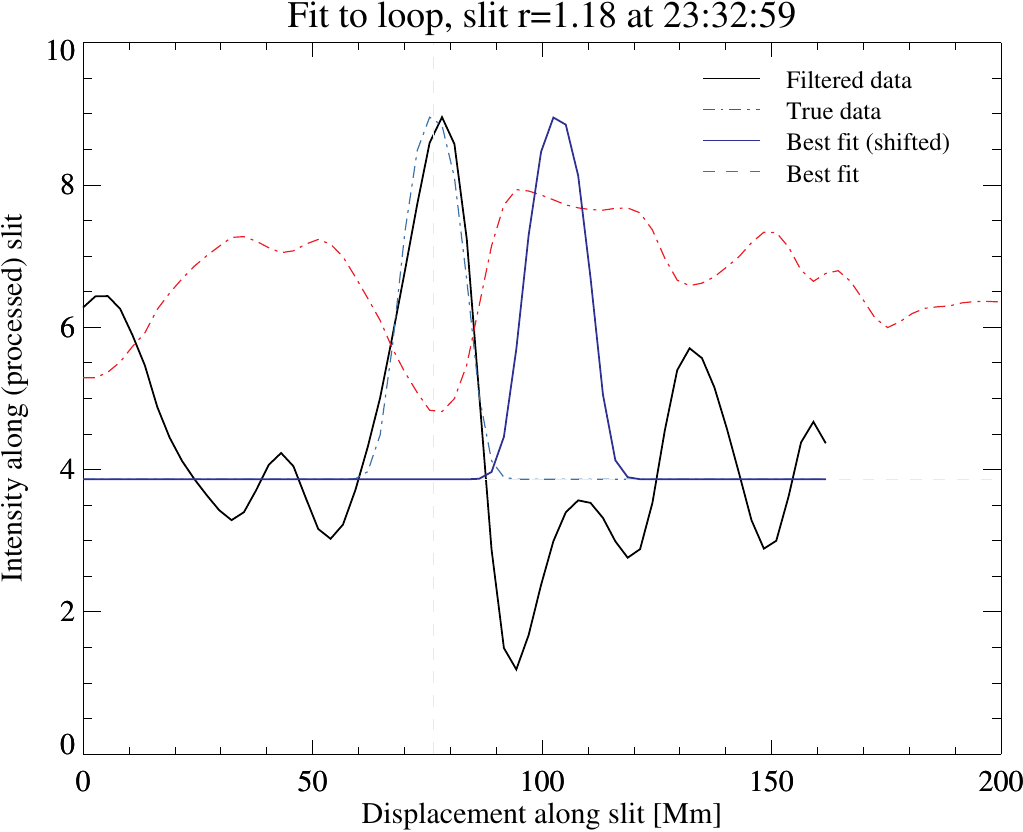}
    %\end{interactive}
    \caption{Plot showing how the fitted density enhancement (blue) relates to the original data (red) and the processed time series (black) on which it was calculated. 
        The processed data has been inverted and an edge filter (in the spatial direction) applied.
        The fitting to the well-contrasted trough (now peak) is shown in dashed blue. 
        The shift from dashed blue line to solid blue line has accounted for the inversion and lies within the region of high intensity (the inhomogeneity of interest).
        %The fit for each time step between 23:27 and 23:47 is available as an animation.
        \label{fig:QCshift}}
\end{figure}

We focus on the time-distance map for the azimuthal slit at $1.18$~\rs, which aligns most closely with the loop apex and therefore shows the clearest transverse motion signature, seen in Figure~\ref{fig:td_maps}. 
Direct fitting to the (raw) intensity map in panel (a) proved unreliable.
Fortunately, the feature seen at $\approx 80$~Mm is a distinct intensity dip, providing excellent contrast to the overlying, less well-defined loop structure at $\approx 100$~Mm. 
We exploit this by fitting the displacement of the loop's `shadow', which exhibits a profile more amenable for our fitting procedure, followed by a shift based on the fitted radius. 
By using a Bayesian framework with SoBaT, we simultaneously optimise for both the displacement and the radius by marginalising the joint posterior distribution for each parameter. 
This accounts for uncertainties in other parameters, allowing the track fitted on the observed `shadow' to be aligned with the density enhancement, seen clearly in Figure~\ref{fig:QCshift}.
The difficulty of fitting a simple profile directly to the broad, non-Gaussian intensity feature at $\sim$100~Mm is apparent when considering the red curve (original data).
Conversely, fitting the inverted and (spatially) edge-filtered data, represented by the black curve, is more reliable and precise as demonstrated by the closeness of the dashed light-blue curve. 
To track the position of the intensity enhancement (loop of interest), we apply a %time-dependent 
shift based on the fitted curve width, with an average shift of approximately 27~Mm. 
% We fit a density radius, and epsilon: how sharply contrasted the loop density at the boundary.
% Note loop's inhomogeneous boundary layer width l = eps*rad.
% Ideally we would want loop width = rad(1+eps) since epsilon is usually ~1, then double. But if epsilon small, then loop width = rad.
% Since epsilon fluctuates, found it best to do some averaging and double (to move across one diameter), so used SHIFT = average[rad*(2 + avg(eps))]. 
This results in the solid blue curve in Figure~\ref{fig:QCshift}, from which we extract the peak position, and connecting each peak in time results in the yellow curves in Figure~\ref{fig:td_maps}.

The derivative of the displacement in the plane of sky is found using a Savitsky-Golay filter of the first order, and we call this the plane-of-sky velocity, $v_{POS}$.
Due to the azimuthal direction of the originating slit, the measured spatial shift and $v_\text{POS}$ are the displacement and velocity in the direction parallel to the surface projected onto the plane of sky.

\subsection{Line of sight velocity}
\label{subsec:LOS_velocity}
\begin{figure}
        \centering
        \includegraphics[width=\linewidth]{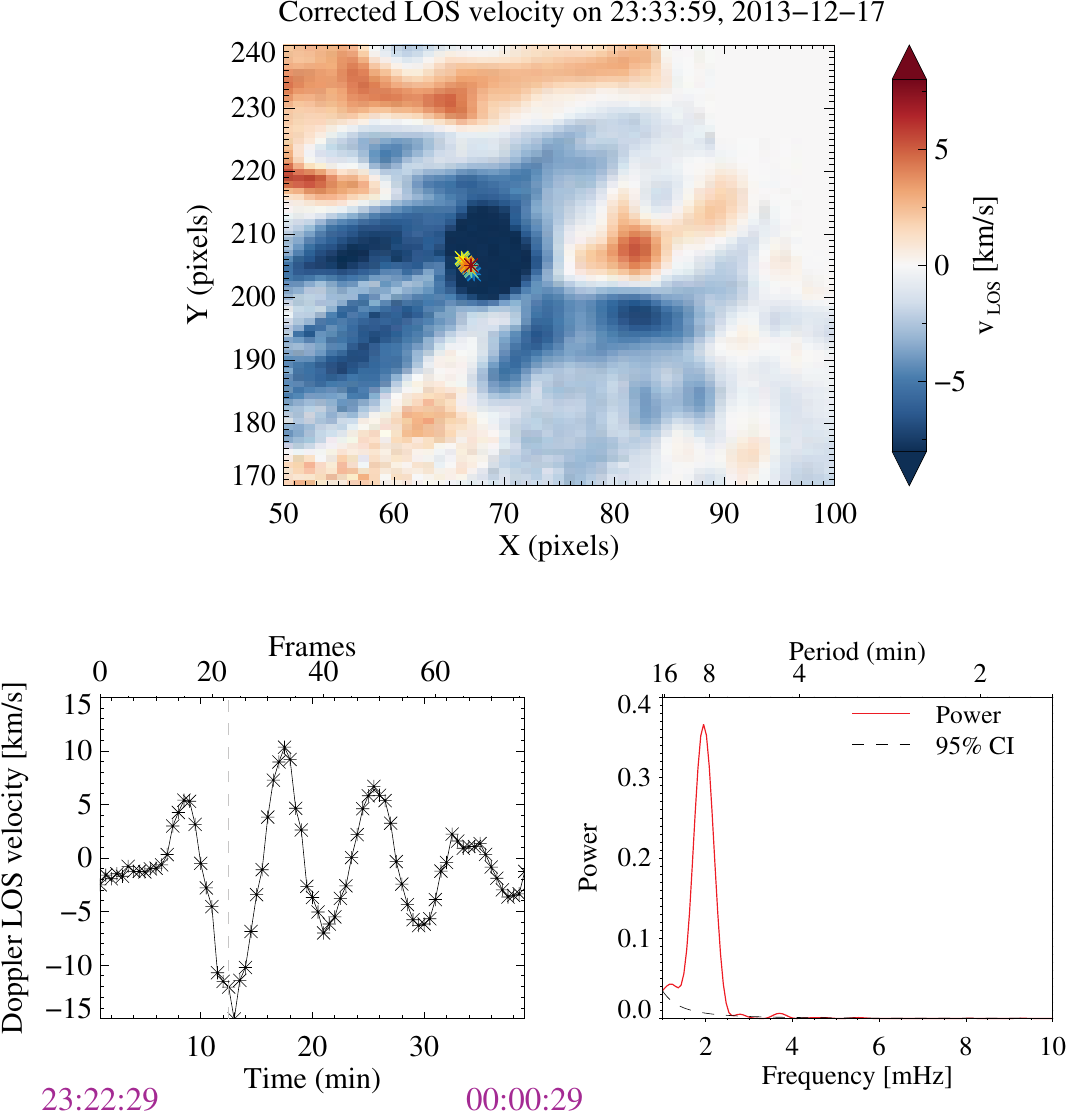}
        \caption{
        The Doppler velocity at the loop apex is tracked as it moves. 
        [Top] shows the location of the points from which the line-of-sight velocity $v_{LOS}$ is extracted. At this scale, these points overlap.
        [Bottom left] the time series $v_{LOS}$ starting from the time of eruption revealing a clearly sinusoidal signal. 
        The dashed vertical line indicates the snapshot displayed above, chosen to be at a velocity oscillation maximum.       
        [Bottom right] the power spectrum of the extracted velocity, with a peak at 2.1 mHz.
        }
        \label{fig:extracted_velocity}
\end{figure}

Through precise tracking of the loop apex's trajectory on a time-distance map and mapping into the CoMP field of view, we determine the line-of-sight Doppler velocity measured at the apex at each moment in time as it moves through space.
The resulting Doppler velocity time series may be seen in Figure~\ref{fig:extracted_velocity}.

As a comparison, spectroscopic analysis of Solar Orbiter STIX data and the Hindoe/EIS spectrograph found the three-dimensional reconstructed plasma flow velocities in steady, unperturbed coronal loops to be of the order $\leq$\qty{5}{\km\per\s} and under \citep[c.f. table 3,][]{Podladchikova2021}. 
The measured Doppler velocity amplitude for this oscillating loop comfortably exceeds this rest value, and matches well the estimated velocity amplitudes of up to \qty{16}{\km\per\s} reported in \citet{Long2017} for a similarly excited loop. 

\section{Results}
\subsection{Oscillation parameters}
The displacement was modelled using $10^5$ Markov Chain Monte Carlo (MCMC) samples, fitting a sinusoidal function with Gaussian damping to represent the observed impulsive excitation. 
The functional form fitted and the best-fitting parameters of initial (projected) displacement amplitude ($-7.0^{+1.3}_{-1.4}$~Mm),
period ($8.9^{+0.5}_{-0.5}$~minutes), 
quality factor ($1.9^{+0.5}_{-0.3}$)
and offset ($109^{+0.6}_{-0.5}$~Mm)
are presented in equation~\ref{eq:displacement}.
%% Displacement result
\begin{equation}
    %  \begin{equation*}
    %     s = P[0] * \sin\left(\frac{2\pi}{P[1]} t \right) * \exp\left(\frac{-t^2}{(2*(P[2]*P[1])^2)}\right) -P[3] \quad \text{[km/s]} \ ,
    % \end{equation*} 
    \begin{aligned}
    s(t) = & -7.0^{+1.3}_{-1.4}~\text{\footnotesize[km/s]} & \text{\footnotesize (amplitude)}\\
            & \times\sin\ \left(\frac{2\pi}{8.9^{+0.5}_{-0.5}~\text{\footnotesize[min]}} t \right) & \text{\footnotesize (period)}\\
            &\times \exp\left(\frac{-t^2}{(2\times(1.9^{+0.5}_{-0.3}\times 8.9^{+0.5}_{-0.5})^2)}\right)  & \text{\footnotesize (damping)} \\
            & + 109^{+0.6}_{-0.5} \quad \text{\footnotesize[Mm]}  & \text{\footnotesize (offset)}\ .
\end{aligned} \label{eq:displacement}
\end{equation}
%The period was determined to be $8.9^{+0.1}_{-0.1}$ minutes based on the posterior distribution. However, accounting for the 30-second sampling cadence; the limited observation of approximately three cycles; and the quality factor of the oscillation being approximately 2, a more realistic uncertainty is estimated to be $\pm 0.5$ minutes. The initial amplitude of the (projected) displacement was found to be $7.0^{+1.3}_{-1.4}$~Mm.

Similarly, the Doppler velocity, $v_{LOS}$, is modelled using a Gaussian damped cosine with the same framework, whose functional form and best-fitting parameters are presented in equation~\ref{eq:vlos}, %
yielding an initial velocity amplitude of $10.9^{+1.6}_{-1.7}$ \unit{km/s}; a derived period of $8.5^{+0.5}_{-0.5}$~minutes with a quality factor just exceeding 2.
 %% Line of sight velocity result
\begin{equation}
    %  \begin{equation}
    %     v_{LOS} = P[0] * \cos\left(\frac{2\pi}{P[1]} t \right) * \exp\left(\frac{-t^2}{(2*(P[2]*P[1])^2)}\right) -P[3] \quad \text{[km/s]} \ ,
    % \end{equation} 
\begin{aligned}
v_{LOS} =  & 10.9^{+1.6}_{-1.7}~\text{\footnotesize[km/s]} & \text{\footnotesize (amplitude)} \\ 
            & \times \cos\left(\frac{2\pi}{8.5^{+0.5}_{-0.5}~\text{\footnotesize[min]}} t \right) & \text{\footnotesize (period)} \\
            &\times \exp\left(\frac{-t^2}{(2\times(2.7^{+0.6}_{-0.4}\times 8.5^{+0.5}_{-0.5})^2)}\right) & \text{\footnotesize (damping)} \\
            & -1.4^{+0.5}_{-0.5}~\text{\footnotesize[km/s]} & \text{\footnotesize (offset)} .
\end{aligned} \label{eq:vlos}
\end{equation}
There are minor discrepancies in quality factor and period because of the short signal length, slight differences in the time series used for the respective fits, and the modelling of the same oscillation from different perspectives.

\subsection{Comparison between Doppler velocity and displacement}
\label{subsec:comparison}
\begin{figure}
        \centering\includegraphics[width=\linewidth]{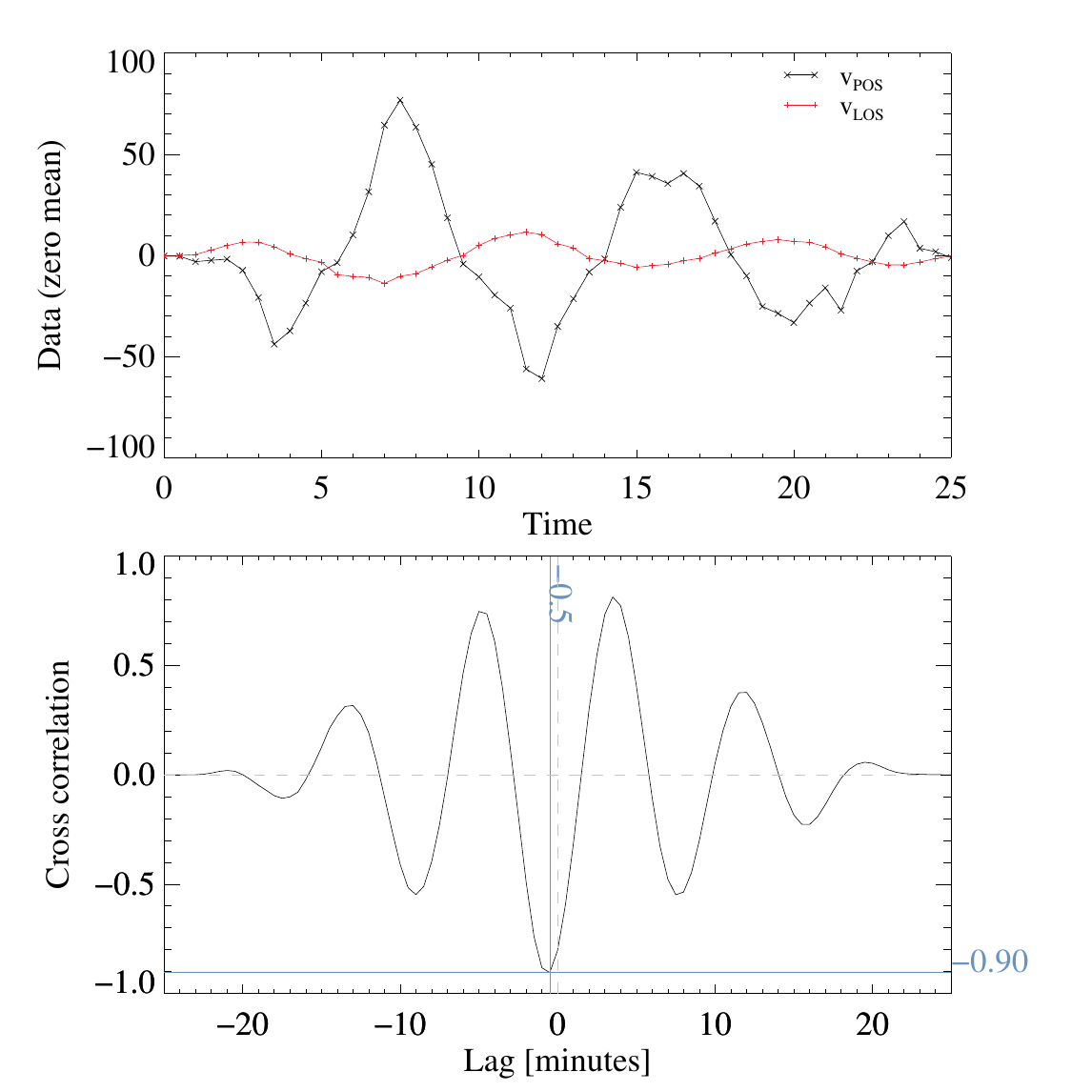}
        \caption{Comparison of line-of-sight Doppler velocity ($v_{LOS}$, red) with the transverse velocity in the plane of sky ($v_{POS}$, black), derived from the derivative of displacement.
        [Bottom] cross correlation of $v_{LOS}$ against $v_{POS}$.
        Clearly the two velocity curves are in antiphase.
        The clear monochromatic sinusoid shows there is a common periodicity of $\sim$8.6 minutes, and there is a very strong correlation coefficient value of 0.89. 
    }
    \label{fig:cc}
\end{figure}

The impulsively driven kink mode oscillation is pronounced in both the Doppler line-of-sight velocity, $v_{LOS}$, and the azimuthal displacement in the plane-of-sky (Figure~\ref{fig:cc}).
Both velocities exhibit a monochromatic oscillation with a common periodicity of $8.5^{+0.5}_{-0.5}$~minutes. 
The derivative of azimuthal displacement, $v_{POS}$, is nearly in anti-phase with $v_{LOS}$, as indicated by a strong correlation coefficient of $-0.89$. %at a lag of one time step (30 seconds).
Note that the sign of $v_{POS}$ depends on the (arbitrary) choice of azimuthal slit direction; if the slit was chosen to be oriented in the opposite direction (e.g., north-to-south, not south-to-north), the sign of $v_{POS}$ would reverse, resulting in in-phase velocities (positive cross-correlation at zero lag).
The dynamics remain consistent: $v_{POS}$ and $v_{LOS}$ reach their maximal magnitudes simultaneously, and both velocities are (approximately) zero when the loop reaches its maximum displacement from equilibrium. 
Noting that radial displacement was found to be negligible, this behaviour confirms that the oscillation is primarily confined to a single plane and symmetric, as a circular or elliptical polarisation would result in a phase difference of approximately \qty{90}{\degree}. 
That is to say, the phase relationship between $v_{POS}$ and $v_{LOS}$ indicate the loop is linearly polarised.
% Depending on the viewing angle, this plane might have components both in the plane of the sky and along the line of sight, although the cross correlation alone is insufficient to reveal the plane of motion.

% Cross correlation

\begin{figure}
    \begin{center}
    	% Span page using figure* and width=0.5\textwidth
        \resizebox{\hsize}{!}{\includegraphics{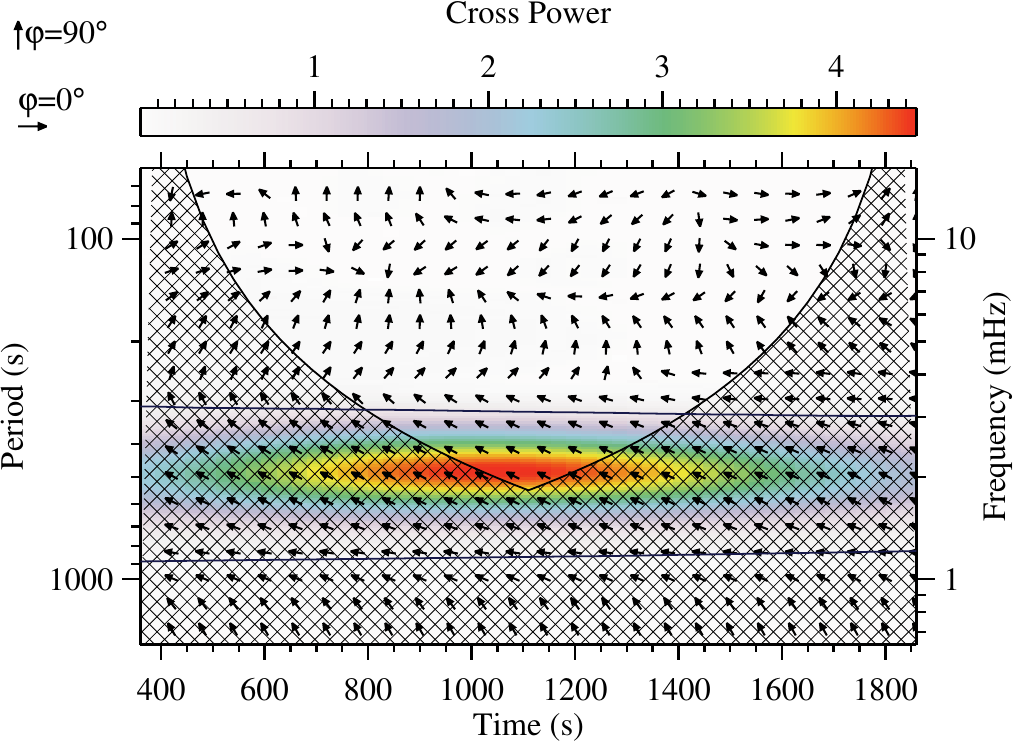}}
        \resizebox{\hsize}{!}{\includegraphics{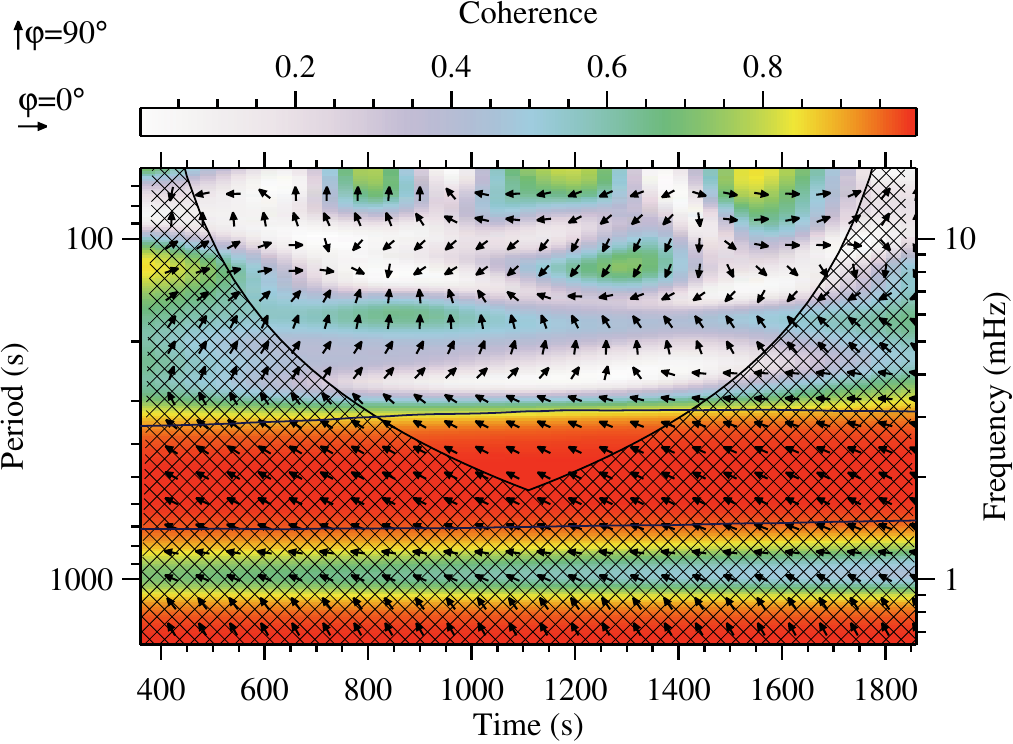}}
	\end{center}
	\caption{Wavelet cross-spectrum and coherence calculated between the plane of sky velocity and the line of sight velocity.
    The phase angle is depicted by arrows where pointing directly left means at 180\unit{\degree} angle.
    The grey hashed lines indicate the cone of influence. 
    The blue contours show the 95\% confidence level calculated with a significance test using 1000 random permutations. 
	}
	\label{fig:crossspec_coherence}        
\end{figure}

The Morlet wavelet cross-spectrum and coherence plots, shown in Figure~\ref{fig:crossspec_coherence} confirm these results.
The cross-power spectrum reveals a single dominant periodicity of 515 seconds, which emerges shortly after the eruption enters the CoMP field of view (at approximately 390 seconds) and subsequently exhibits damping. 
The coherence at this frequency is exceptionally high, indicating a strong and consistent relationship between the signals over time. 
The phase arrows overlaid on both spectra confirm a steady phase relationship between the signals, at approximately \qty{150}{\degree}.

In contrast to the cross-correlation that provides a time-averaged value, the wavelet cross-spectrum provides a time-resolved view of the phase relationship, allowing us to track how the phase between line-of-sight and plane-of-sky velocities evolve over the duration of the oscillation. 
A deviation of the observed phase difference in the cross-spectrum from \qty{180}{\degree} (or \qty{0}{\degree} depending on the convention of $v_{POS}$) suggests the influence of resonant absorption. 
As energy is transferred from the global kink mode to the local Alfv\'en continuum modes, the flux tube's motion changes from a purely transverse oscillation to one with an increasingly significant azimuthal component, as shown analytically by \citet{Goossens2014}  % In the words of \citet[section 6][]{Goossens2014}, since we can decompose a (their equation 30) 
and numerically by \citet{Antolin2017}. % Though in this case they compare POS motion, just take a derivative.
This process introduces an additional phase offset between the azimuthal and radial velocities, which is then projected onto the plane-of-sky and line-of-sight, resulting in the observed phase difference. 
This provides further evidence that resonant absorption may be the dominant damping mechanism in coronal loop kink oscillations, reinforcing previous results that showed how the quality factors for multiple harmonics agree \citep[as expected from resonant absorption;][]{Duckenfield_3rd}, which is further substantiated by theoretical considerations \citep[e.g.,][]{Ruderman2009_KinkReview, Guo2020_elliptical1,Morton2021_weakdamping}.

% Cross spectrum + coherence.

\subsection{Interpretation of hodogram: polarisation}
\label{subsec:obs_polarisation}
% Hodogram
\begin{figure}
    %\begin{interactive}{animation}{fig8.mp4}
    \gridline{\fig{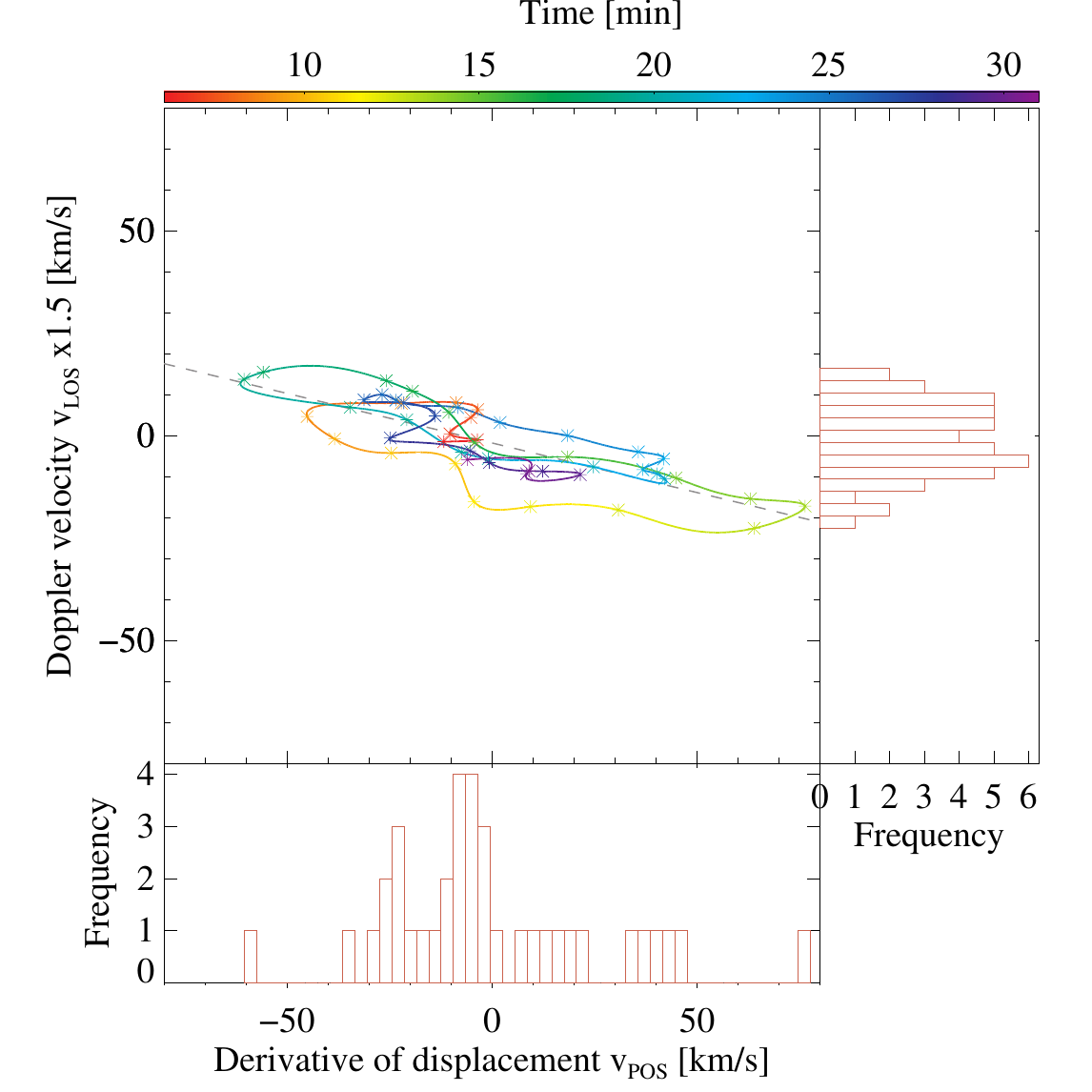}{\linewidth}{(a) Hodogram}}
     \gridline{\fig{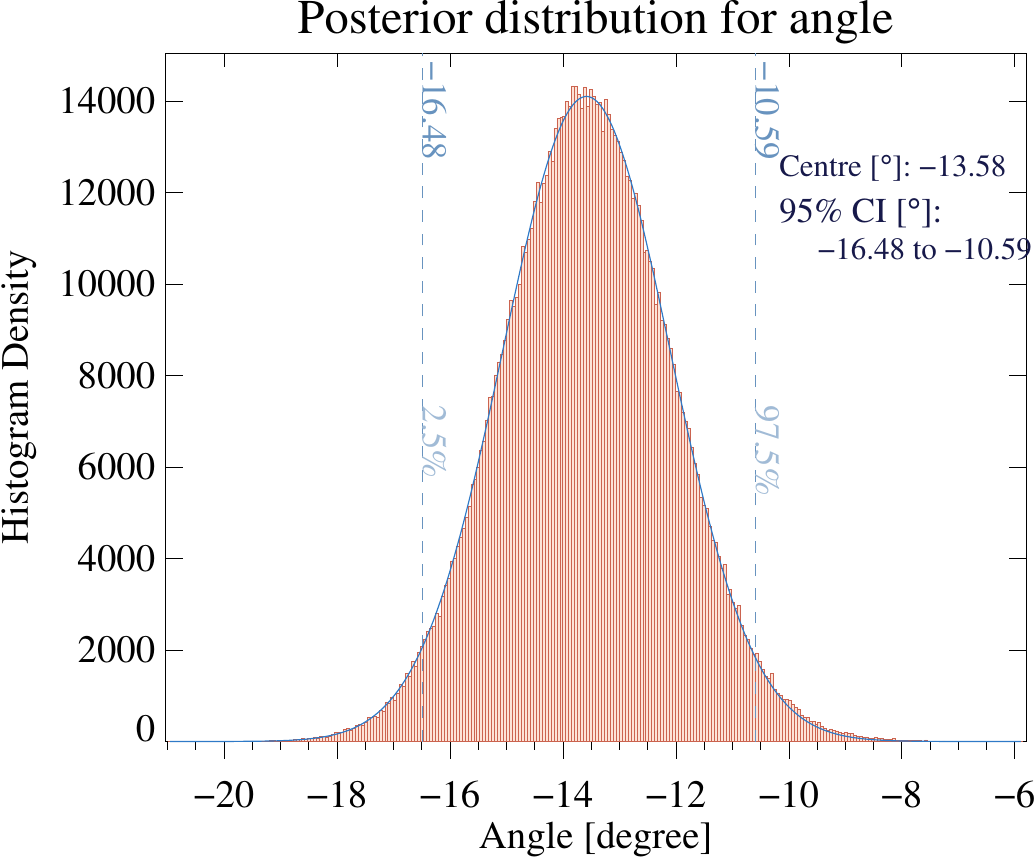}{0.5\linewidth}{(b) Posterior for angle}
    \fig{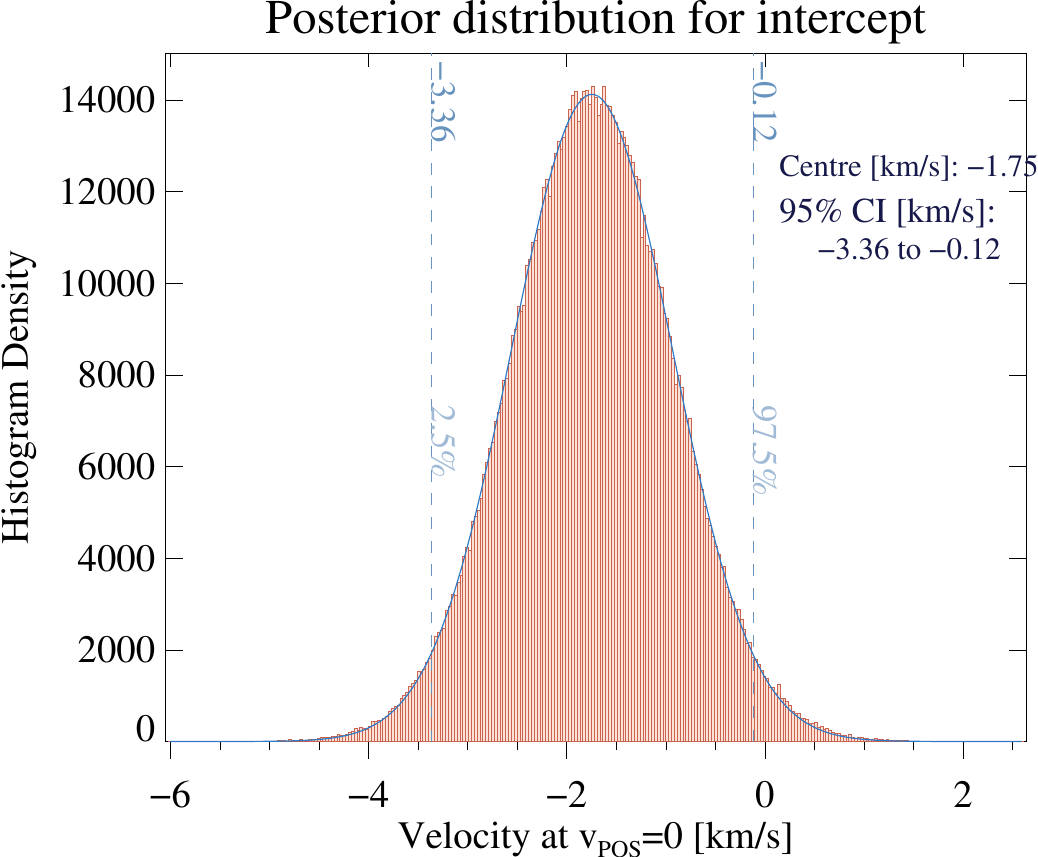}{0.5\linewidth}{(c) Posterior for intercept} }
    %\end{interactive}
    \caption{Hodogram showing the evolution of the velocity vector with time.
    The plane-of-sky velocity derived from the derivative of the fitted displacement is shown along the x-axis, and the Doppler (line-of-sight) velocity extracted by tracking the feature, scaled by $\times$1.5, is shown along the y-axis.
    %The dominant component of the velocity is in the plane of sky, seen as $v_{POS}$ being of greater amplitude than its counterpart.
    The loop's velocity vector, as visualized in the velocity-space hodogram, oscillates along a line orientated at -13.6\unit{\degree} relative to the plane of sky.
    %This oscillation can be seen in the accompanying animation of the hodogram.
    The bottom panels show normalised histograms approximating the marginalized posterior distributions of the (b) angle and (c) intercept obtained from $10^5$ MCMC samples using the model $v_{LOS} = \tan(\text{angle})*v_{POS} + \text{intercept}$.
    }
    \label{fig:hodogram}
\end{figure}
Hodograms provide a powerful visualisation of oscillatory motion by plotting velocity components against each other, revealing the polarisation state and underlying geometry of the oscillation \citep[e.g.,][]{Zhong2023_polarisation, Bate2024_polarisedfibril}. % Also Brannon2015
Plotting the velocity vector for this loop oscillation $(v_{POS}, v_{LOS})$ with time (Figure~\ref{fig:hodogram}) reveals a remarkably linear phase portrait, even without any filtering, and confirms that the loop oscillation is linearly polarised. 
The angle of the hodogram, assuming negligible radial velocity, indicates the orientation of the oscillation in the plane of motion, specifically that the loop sways towards the Sun's southern pole as Earth sees it ($v_{\text{POS}} < 0$~km{\,}s$^{-1}$) while simultaneously moving away from the observer ($v_{\text{LOS}} > 0$~km{\,}s$^{-1}$). 
Subsequently, as the loop returns back to its equilibrium point it moves azimuthally towards the solar north and towards from the observer.
The shallow angle tells us that the plane of oscillation is closely aligned to the plane of sky. 

As mentioned, assuming zero radial displacement and fitting a line of best fit to the hodogram of $v_{LOS}$ versus $v_{POS}$ provides a direct measure of the orientation of the oscillation plane, as the velocities are constrained to the azimuthal and line-of-sight directions. 
We correct for the known underestimation of Doppler velocities by CoMP through multiplying $v_{LOS}$ by a factor of $\times1.5$ based on \citet{Lee2021_CoMP}, and discussed in \ref{subsec:CoMP}. %\ref{subsec:underestimation}.
Using the corrected $v_{LOS}$, the line of best fit for the hodogram in Figure~\ref{fig:hodogram} is computed with the Bayesian methodology of SoBaT \citep{Anfinogentov2020_SOBAT}.
For the angle variable, a uniform prior $\mathcal{U}(-\qty{180}{\degree},\qty{180}{\degree})$ is used, since we have no expectation for a specific polarisation plane. 
For the intercept variable, a normal prior $\sim \mathcal{N}(0,\sigma_{v})$ 
was used since we expect no bulk motion besides our uncertainty in both velocities, which we (significantly) overestimate using the standard deviation of both $v_{LOS}$ and $v_{POS}$, $\sigma_v \approx \qty{22}{\km\per\second}$.
The line of best fit of the hodogram of Figure~\ref{fig:hodogram} is found to be:
\begin{equation*}
    v_{LOS} = \tan\left(-13.6^{+2.9}_{-3.0} \unit{\degree} \right)*v_{POS} -1.77^{+1.59}_{-1.65} \quad \text{[km/s]} \ ,
\end{equation*} 
where the error ranges given are for the 95\% confidence interval.
Testing showed the results were insensitive to the choice of priors.
The bottom panels (b) and (c) of Figure~\ref{fig:hodogram} show the well-constrained marginalised posterior distributions for the angle and intercept, estimated by $10^5$ MCMC samples. 

\section{Discussion}
\label{sec:conclusions}
Reiterating the assumption of zero radial displacement, its phase behaviour suggests the kink oscillation in this study is linearly polarised and oscillating in a plane approximately \qty{14}{\degree} tilted from the plane of sky, such that the loop apex alternates between moving away from Earth \& azimuthally southward, and moving towards Earth \& azimuthally northward.
% Phase fine regardless of x1.5
Note that the calculations of cross correlation, cross spectrum, and coherence are primarily sensitive to the phase relationship between $v_{LOS}$ and $v_{POS}$ and are thus unaffected by any underestimation of velocity amplitudes. 
No matter the level to which $v_{LOS}$ is underestimated, the linear polarisation of the kink oscillation remains, though the resultant angle of the plane of oscillation would be different. 
The same calculation for $v_{LOS}$ without scaling yields an angle of $9.2^{+2.1}_{-2.0}$\unit{\degree} and an intercept of $-1.2^{+1.1}_{-1.1}$ \unit{\km\per\second}.  

% Discussion?
The non-zero intercept of the line calculated for the hodogram Figure~\ref{fig:hodogram} may imply some bulk motion of the loop, in this case towards the observer at approximately \qty{2}{\km \per \second}. %  or some short-lived restructuring of the active region from the eruption.
Note that the CoMP Level 2 FITS files have been corrected for solar rotation effects, at least partially.
A smoothed polynomial fit, combined with a zero-median assumption for each frame, was used to adjust the Doppler velocity measurements, accounting for the blueshift at the east limb and redshift at the west limb \citep{Tomczyk2021_uCoMP}.
Nonetheless this small bulk velocity could be a residual from the solar rotation.

\subsection{Line widths}
\label{subsec:line_widths}
% Line width
\begin{figure}
        \centering
        \includegraphics[width=\linewidth]{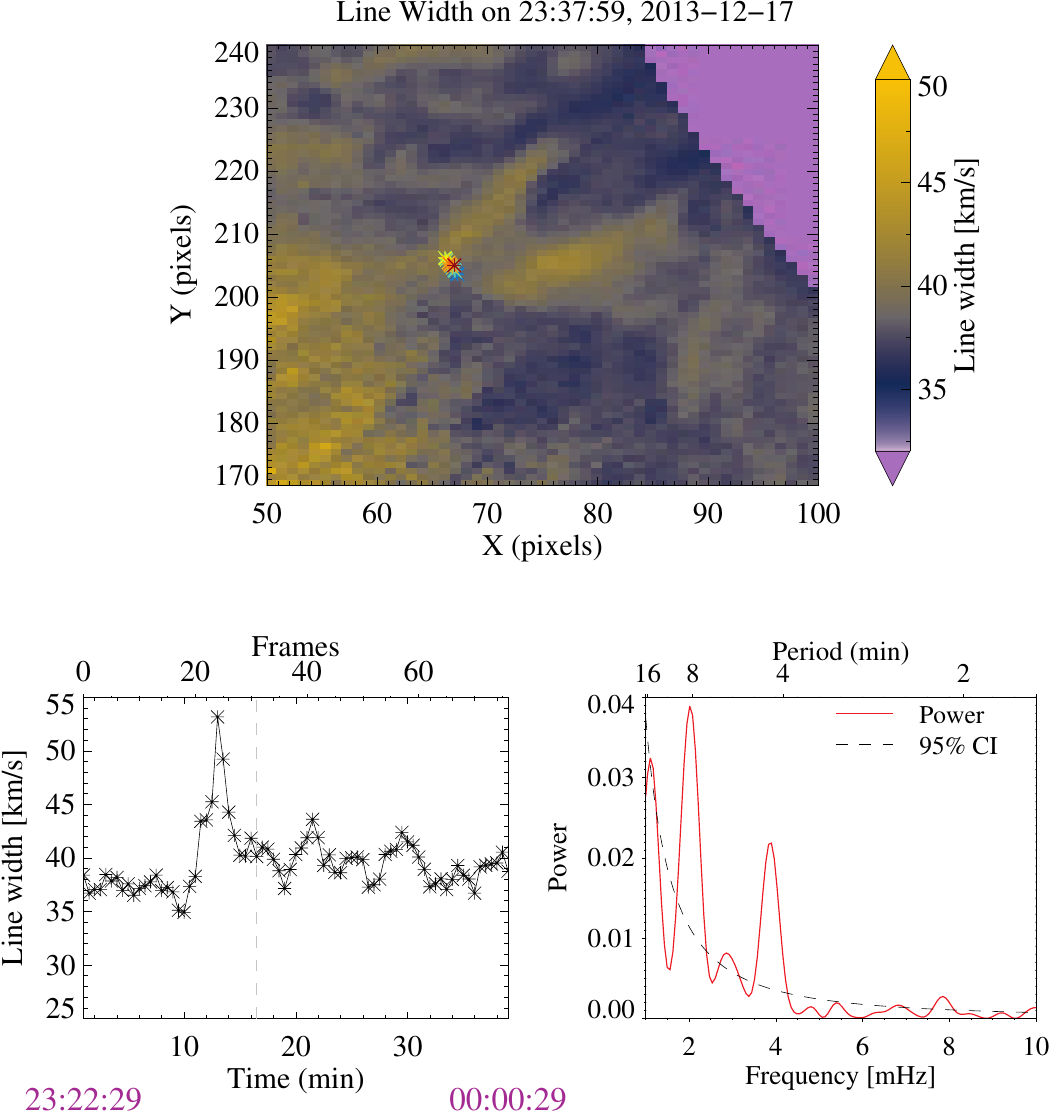}
        \caption{
        The line width at the loop apex (1.18~\rs) is tracked as it moves. 
        [Top] shows the location of the points from which the line width is extracted. 
        %At this scale, these points overlap, though the loop is clearly visible.
        [Bottom left] the time series of line width at loop apex.
        The vertical dashed line indicates the snapshot displayed above, chosen to best highlight the loop structure.      
        [Bottom right] the power spectrum of the extracted line widths, showing two peaks (2.0 and 3.9 mHz) above the 95\% confidence level. 
        }
        \label{fig:extracted_linewidth}
\end{figure}
As with the Doppler velocities, by precisely tracking the loop as it moves we can extract the line width from its apex. 
Referring to Figure~\ref{fig:extracted_linewidth}, there is a clear periodic increase in line width as the loop oscillates.
Curiously the line widths appear broadened throughout the entire loop, giving credibility.
The line width enhancement is at a maximum when $v_{LOS}$ is at its most negative (blue-shifted) and $v_{POS}$ is at its greatest (i.e., when the inhomogeneity is passing through its equilibrium point). % confirmed by the cross correlation of displacement vs width. 
The dominant period matches that of the kink oscillation, 8~minutes, however a peak at the double frequency is also above the 95\% confidence level. It is expected that a peak at twice the frequency is related to two wave maxima being present during each complete oscillation. 
Previously, this type of behaviour has been attributed to unresolved line width fluctuations synonymous with torsional Alfv{\'{e}}n waves \citep[e.g.,][]{2009Sci...323.1582J, 2013SSRv..175....1M}.
Furthermore, \citet[][]{Antolin2017} attributed this to large shear flows at the boundary layer of the loop, which subsequently induce eddies and instabilities that appear as spectral line broadening when unresolved.
As these shear flows are greatest as the loop reach their maximum speed regardless of the direction of motion, the period of line enhancements is expected to be doubled.
% For CoMP there is a known correlation between Doppler velocity and line width \citet{Lee2021_CoMP}, so it may be that there is some cross-talk happening.
% Furthermore CoMP is found to underestimate non-thermal line widths in bright regions by 20\% to 40\%, even more so during a flaring event.  
% Furthermore, a superposition of multiple oscillating coronal loops or structures, nearly in phase but differing slightly in properties such as loop length and initial displacement timing, could potentially contribute to an apparent line broadening.
There is also a remote possibility that the line width increase is due to the thermalisation of the plasma, increasing the thermal width.
The exact mechanism is unclear from this observation alone.
Nonetheless, the detection of a double frequency oscillation in line width for an isolated kink oscillation is, to the author's knowledge, the first of its kind, and will be followed up in a future publication.

\subsection{Outlook}
\label{subsec:Outlook}
The approach outlined in this work — combining Doppler velocities and plane-of-sky tracking to infer the polarisation of transverse motions in solar structures — can be achieved using a single instrument, has the potential to revolutionise coronal seismology.
By systematically supplementing observations of coronal waves with constraints on their polarisation, we can better estimate their velocity amplitudes and absolute displacements, which will help seismologists derive more accurate estimates of energy fluxes and deposition rates. 
Ultimately, quantifying how much energy waves transport, and how they contribute to coronal heating, will bring us closer to resolving the coronal heating paradox.  
Additionally, observing a wave's polarisation and how it may change over time provides additional information observers can use to better understand the underlying magnetic geometry and dynamics.

In the general case, the inference of polarisation may require an analysis of all three velocity components (both plane-of-sky displacements and Doppler velocity) and their phase relationship, but the principle is unchanged. 
As demonstrated here the Coronal Multi-channel Polarimeter (CoMP) is well-suited to this technique, and by providing continuous, long-duration observations of the entire corona it enables comprehensive studies of wave polarisation (as well as propagation) through the solar corona.
Excitingly, the instrument has since been upgraded (now referred to as uCoMP) to a larger field of view (previously 1.05 -- 1.3 \rs, now 1.03 -- 1.95 \rs);
wider spectral range (\qtyrange{1074}{1083}{\angstrom}, now \qtyrange{530}{1083}{\angstrom});
and improved spatial resolution (4.5~arcsec/pixel, now 3~arcsec/pixel), taking daily scientific observations from mid-2021.\footnote{Note an eruption of nearby volcano Mauna Loa on November 28\textsuperscript{th} 2022 closed the observatory%\href{https://www2.hao.ucar.edu/mlso/2022-eruption-of-mauna-loa}{, see their page for updates}
. Repairs are due to be completed by the end of 2024.} 
This upgrade will allow more precise tracking of transverse motions across a wider range of coronal structures, while the expanded spectral coverage will enable simultaneous analysis of additional lines, improving the accuracy and applicability of the polarisation inference technique demonstrated in this paper, and something we aim to explore in a future paper.

The upcoming Multi-slit Solar Explorer \citep[MUSE;][]{DePontieu2020_MUSEapproach} mission, designed to deliver high-resolution spectroscopy and imaging of the solar corona, will also be able to employ this technique, although there will be some differences. 
While MUSE's smaller field of view and shorter observation windows limit it to case-by-case studies, its ability to resolve fine spatial and temporal details, along with complementary observations across multiple spectral lines, offers great potential for advancing our understanding of wave dynamics, energy transport, and dissipation.

\subsection{Conclusion}
\label{subsec:conclusions}
This paper introduces a novel method in coronal seismology that combines the tracking of oscillations in the imaged plane with Doppler line-of-sight velocities. 
By fitting the best-fit angle to the hodogram (or phase portrait) of the oscillation, the polarisation can be inferred.
For this example using CoMP data, a kink oscillation excited by a small coronal eruption was found to be linearly polarised, oscillating in an azimuthal direction in a plane roughly \qty{14}{\degree} inclined from the plane of sky.
The wavelet cross-spectrum showed a small deviation of the phase shift between the Doppler velocity and projected plane-of-sky velocities as expected from resonant absorption; the process which, according to theoretical/analytical works, is an efficient mechanism for damping kink waves by transferring their energy into unresolved azimuthal Alfv\'en waves, which can then dissipate their energy into heating.
An intriguing hint of double frequency periodicity was detected in the line width data, which adds to the suggestion of instabilities at the oscillating loop boundary. 
We anticipate the broad applicability of this hodogram tool to the enhanced observations of the upgraded uCoMP system. 
The improved spatial and temporal resolution of uCoMP, coupled with its continuous monitoring of the entire corona, increases the likelihood of capturing fortuitous events for detailed study. This enables the widespread study of kink oscillations using a single instrument.
Stereoscopy is a timely topic, especially with the wealth of solar observations from missions like Solar Orbiter and MUSE. Hodograms offer a complementary tool for analysing these stereoscopic datasets.
This technique has potential applications to other aspects of coronal science. 
The relevant codes for the analysis and techniques used in this paper are available through the \textit{Waves in the Lower Solar Atmosphere} (WaLSA) coding repository, \textbf{WaLSAtools}\footnote{WaLSAtools website: \url{https://WaLSA.tools}} \citep{2025NRMP....wave.analysis.tools}. 
The data used in this study can be provided upon reasonable request to the corresponding author.

\begin{acknowledgments}
TJD and DBJ thank the support of the Leverhulme Trust via the Research Project Grant RPG-2019-371. 
DBJ and SJ wish to thank the UK Science and Technology Facilities Council (STFC) for the consolidated grants ST/T00021X/1 and ST/X000923/1. 
DBJ also acknowledges funding from the UK Space Agency via the National Space Technology Programme (grant SSc-009).
RJM is supported by a UKRI Future Leader Fellowship (RiPSAW—MR/T019891/1). 
Finally, we wish to acknowledge scientific discussions with the Waves in the Lower Solar Atmosphere (WaLSA; \href{https://www.WaLSA.team}{https://www.WaLSA.team}) team, which has been supported by the Research Council of Norway (project no. 262622), The Royal Society \citep[award no. Hooke18b/SCTM;][]{2021RSPTA.37900169J}, and the International Space Science Institute (ISSI Team~502). 
%RM
%This work was supported by the the British Council via the Institutional Links Programme (Project 277352569 `Seismology of Solar Coronal Active Regions'), the STFC consolidated grant ST/L000733/1.
%DJP was supported by the GOA-2015-014 (KU Leuven) and the European Research Council (ERC) under the European Union's Horizon 2020 research and innovation programme (grant agreement No 724326). 
%CRG was supported by ERC Synergy grant WHOLE SUN 810218.
\end{acknowledgments}

\vspace{5mm}
\facilities{CoMP, SDO(AIA)}

\bibliography{CoMP_polarisations}{}
\bibliographystyle{aasjournal}

\end{document}